\documentclass[pra,twocolumn,a4paper,superscriptaddress]{revtex4}
\usepackage{amsmath,amssymb,graphicx,color,dsfont}

\newcommand{\traceSubInline}[2]{\textrm{Tr}_{#1} \! (#2)}

\newcommand{\abs}[1]{\left|#1\right|}

\newcommand{\integ}[1]{\ensuremath{\int \!\! \mathrm{d}#1 \,}}
\newcommand{\integlim}[3]{\ensuremath{\int_{#1}^{#2} \!\!\! \mathrm{d}#3 \,}}

\newcommand{\hc}{\textrm{H.c.}}

\newcommand{\mean}[1]{\ensuremath{\langle #1 \rangle}}

\newcommand{\prob}[1]{\textrm{Pr} \! \left(#1\right)}

\newcommand{\nbar}{\ensuremath{\bar{n}}}

\newcommand{\omegam}{\ensuremath{\omega_M}}

\newcommand{\detune}[1]{\ensuremath{\Delta_{#1}^{\phantom{\dagger}}}}

\newcommand{\bsParam}{\ensuremath{{\textstyle \frac{\theta}{2}}}}

\newcommand{\bra}[1]{\ensuremath{\left\langle #1 \right|}}
\newcommand{\ket}[1]{\ensuremath{\left| #1 \right\rangle}}

\newcommand{\UpsH}{\ensuremath{\Upsilon_{\! h}^{\phantom{\dagger}}}}
\newcommand{\UpsHd}{\ensuremath{\Upsilon_{\! h}^\dagger}}

\newcommand{\UpsPerp}{\ensuremath{\Upsilon_{\!\!\perp}^{\phantom{\dagger}}}}

\newcommand{\Ueff}{\ensuremath{U_{\textrm{eff}}}}

\newcommand{\ai}{\ensuremath{a_{\textrm{in}}}}

\renewcommand{\ao}{\ensuremath{a^{\phantom{\dagger}}_{\textrm{out}}}}
\newcommand{\aod}{\ensuremath{a^\dagger_{\textrm{out}}}}

\newcommand{\bzero}{\ensuremath{b^{\phantom{\dagger}}_{0}}}
\newcommand{\bzerod}{\ensuremath{b^{\dagger}_{0}}}

\newcommand{\PM}{\ensuremath{P_M}}

\newcommand{\rhoM}{\ensuremath{\rho_M}}
\newcommand{\rhoMin}{\ensuremath{\rho_M^{\textrm{in}}}}
\newcommand{\rhoMout}{\ensuremath{\rho_M^{\textrm{out}}}}

\newcommand{\jj}[1]{\,{#1}\,}

\newcommand{\etal}{\emph{et al}.}
\newcommand{\PRL}[3]{Phys. Rev. Lett.~\textbf{#1}, #2 (#3)}
\newcommand{\PRA}[3]{Phys. Rev. A~\textbf{#1}, #2 (#3)}
\newcommand{\PRAR}[3]{Phys. Rev. A~\textbf{#1}, #2(R) (#3)} 

\newcommand{\RMP}[3]{Rev. Mod. Phys.~\textbf{#1}, #2 (#3)}
\newcommand{\Nature}[3]{Nature (London)~\textbf{#1}, #2 (#3)}

\newcommand{\NatPhot}[3]{Nature Photonics~\textbf{#1}, #2 (#3)}
\newcommand{\Science}[3]{Science~\textbf{#1}, #2 (#3)}

\newcommand{\APL}[3]{Appl. Phys. Lett.~\textbf{#1}, #2 (#3)}
\newcommand{\EPL}[3]{Europhys. Lett.~\textbf{#1}, #2 (#3)}
\newcommand{\JPhysB}[3]{J. Phys. B \textbf{#1}, #2 (#3)}
\newcommand{\JOptB}[3]{J. Opt. B \textbf{#1}, #2 (#3)}
\newcommand{\NJP}[3]{New J. Phys. \textbf{#1}, #2 (#3)}

\begin{document}

\title{Quantum State Orthogonalization and \\
a Toolset for Quantum Optomechanical Phonon Control}

\author{M.~R.~Vanner}
\affiliation{
\mbox{Vienna Center for Quantum Science and Technology (VCQ) and }\\ 
\mbox{Faculty of Physics, University of Vienna, Boltzmanngasse 5, A-1090 Vienna, Austria}}
\author{M.~Aspelmeyer}
\affiliation{
\mbox{Vienna Center for Quantum Science and Technology (VCQ) and }\\ 
\mbox{Faculty of Physics, University of Vienna, Boltzmanngasse 5, A-1090 Vienna, Austria}}
\author{M.~S.~Kim}
\affiliation{
\mbox{QOLS, Blackett Laboratory, Imperial College London, SW7 2BW, United Kingdom}}

\date{Submitted for review March 16, 2012}

\begin{abstract}
We introduce a method that can orthogonalize any pure continuous variable quantum state, i.e. generate a state $|\psi_\perp\rangle$ from $|\psi\rangle$ where $\langle\psi|\psi_\perp\rangle = 0$, which does not require significant \emph{a priori} knowledge of the input state. We illustrate how to achieve orthogonalization using the Jaynes-Cummings or beam-splitter interaction, which permits realization in a number of systems. Furthermore, we demonstrate how to orthogonalize the motional state of a mechanical oscillator in a cavity optomechanics context by developing a set of coherent phonon level operations. As the mechanical oscillator is a stationary system such operations can be performed at multiple times, providing considerable versatility for quantum state engineering applications. Utilizing this, we additionally introduce a method how to transform any known pure state into any desired target state.
\end{abstract}

\maketitle


A qubit basis formed by a pair of orthogonal quantum states is central to quantum information processing. Currently there is considerable effort towards implementing quantum information processing with two-level systems. For such systems, an intriguing and fundamental fact is that quantum mechanics prohibits the construction of a universal-NOT gate that would produce an orthogonal qubit from any input qubit~\cite{ref:qubitOrthog}. This quantum mechanical property is closely related to the quantum no-cloning theorem~\cite{Scarani2005}, however, faithful cloning can be achieved probabilistically provided that the set of input states is linearly independent~\cite{Duan1998}. Similarly, using such an input set of states, it is possible to construct a probabilistic NOT operation for qubits~\cite{Yan2012}. A qubit basis may, however, also be formed using two orthogonal continuous variable states. Thus far, efforts to construct such a basis have mainly concentrated on using a superposition of coherent states~\cite{Jeong2002}. Also, recently a qubit basis was realized using photon subtraction from squeezed vacuum~\cite{Neergaard-Nielsen2010}.


In this Letter, we introduce a method for quantum state orthogonalization for continuous variable quantum systems. Notably, the method only requires knowing the angle $\vartheta$ made by the state's mean amplitude $\mean{b} = \abs{\mean{b}}e^{i\vartheta}$, where $b$ is the annihilation operator, and hence the scheme is magnitude independent. Furthermore, our method is readily extended to generate an arbitrary superposition of the initial state and an orthogonal counterpart to allow the encoding of quantum information. The orthogonalizer $\UpsPerp \jj{\propto}  b e^{-i\phi} + b^\dagger e^{i\phi}$ is formed by a linear superposition of the bosonic annihilation  and creation operators and generates a state orthogonal to any pure state $\ket{\psi}$ i.e. $\bra{\psi} \UpsPerp \ket{\psi} \jj{=} 0$ when $\phi \jj{=} \vartheta+\pi/2$. Thus, $\UpsPerp$ is a quadrature operator that is perpendicular to $\vartheta$~\cite{note:SpinHalfAnalog}.


The orthogonalizer can be realized with interactions that are available in many physical systems, e.g., to realize $\UpsPerp$ in cavity-quantum-electrodynamics \cite{Kimble1998, Raimond2001}, one prepares an input qubit in the state $A\ket{g} + B\ket{e}$ which then weakly interacts via the Jaynes-Cummings Hamiltonian $H/\hbar = -i\Omega(b \sigma_+ - b^\dagger \sigma_-)$, where $\Omega$ is the coupling rate and $\sigma_{+,-}$ are the raising and lowering operators. A controllably weighted superposition of addition and subtraction is achieved by projective measurement of the qubit onto $B^*\ket{g}\jj{-}A^*\ket{e}$. The measurement operator is then $\Upsilon_{QED} = (\bra{g}B - \bra{e}A)(1-\Omega\tau(b \sigma_+ - b^\dagger \sigma_-))(A\ket{g} + B\ket{e}) = \Omega\tau(A^2b + B^2b^\dagger)$, see Fig.1~(a). With this interaction, optical~\cite{Kimble1998} or microwave~\cite{Raimond2001} fields in a cavity, or the motional state of trapped ions~\cite{Leibfried2003}, can be orthogonalized by appropriately setting $A$ and $B$. Similarly, a pure state of a traveling optical field can be orthogonalized by interaction on a beam-splitter and then measurement of an optical qubit comprising a superposition of zero and one photons \cite{Babichev2003}, see Fig.1~(b). As these interactions are common throughout quantum optics, adaptations of this orthogonalization protocol to other physical systems can be readily achieved. Moreover, a different scheme to perform a superposition of photon subtraction and addition was recently proposed~\cite{LeeNha2010}, which could also be used to realize state orthogonalization. 




The tools we introduce for orthogonalization can also be utilized for quantum state engineering applications. Currently, single-quanta-manipulation techniques performed on traveling light fields~\cite{Kim2008} have prepared superposition states via photon subtraction~\cite{ref:photonSubtraction}, observed the bosonic commutation relation~\cite{ref:bosonComm}, and engineered arbitrary quantum states up to the two-photon level~\cite{Bimbard2010}. Much progress has also been made for arbitrary quantum state preparation of the motion of trapped ions and microwave field states \cite{ref:arbQSP}. As mechanical  elements are now also considered for quantum applications, experimental tools are required for the coherent manipulation of phononic modes. Examples of progress in this direction are the observation of the ground state of motion~\cite{OConnell2010, TeufelGS, Chan2011}, steps towards single-phonon manipulation by coupling to a superconducting phase qubit~\cite{OConnell2010}, strong coupling \cite{ref:strongCoupling}, and mechanical mode thermometry via sideband asymmetry~\cite{Safavi2012}. Also, recently the lattice vibrations of two diamonds were entangled by coherently distributing one quanta across the two vibrational modes~\cite{Lee2011}.

\emph{Coherent phonon manipulation.---}
In this section we demonstrate how to perform an arbitrary coherent superposition of phonon subtraction, addition and the identity operation to a mechanical oscillator using cavity optomechanics. The prototypical optomechanical system is a Fabry-P\'{e}rot cavity where one of the mirrors is sufficiently compliant that the reflection of light can modify the mirror momentum via radiation-pressure. Concurrently, the motion of the moving mirror modulates the optical phase and generates sidebands. To realize phonon subtraction (addition) one can optically drive an optomechanical cavity at the red (blue) sideband and then perform single photon detection on the field scattered onto cavity resonance. Provided that the sidebands are well resolved and the optical phase shifts are small allowing linearization, the red-detuned drive gives rise to a beam-splitter interaction and the blue-detuned drive gives rise to a two-mode-squeezing interaction. This linearization procedure was discussed, for optomechanics, in Ref.~\cite{Zhang2003}, where quantum state transfer between light and mechanics was proposed. Drive on the blue sideband has also been considered for continuous-variable teleportation from light to the mechanics \cite{ref:teleportation}. Some other applications utilizing these sidebands are reviewed in Refs.~\cite{ref:reviews}.

\begin{figure}[t!]
\includegraphics[width=0.8\hsize]{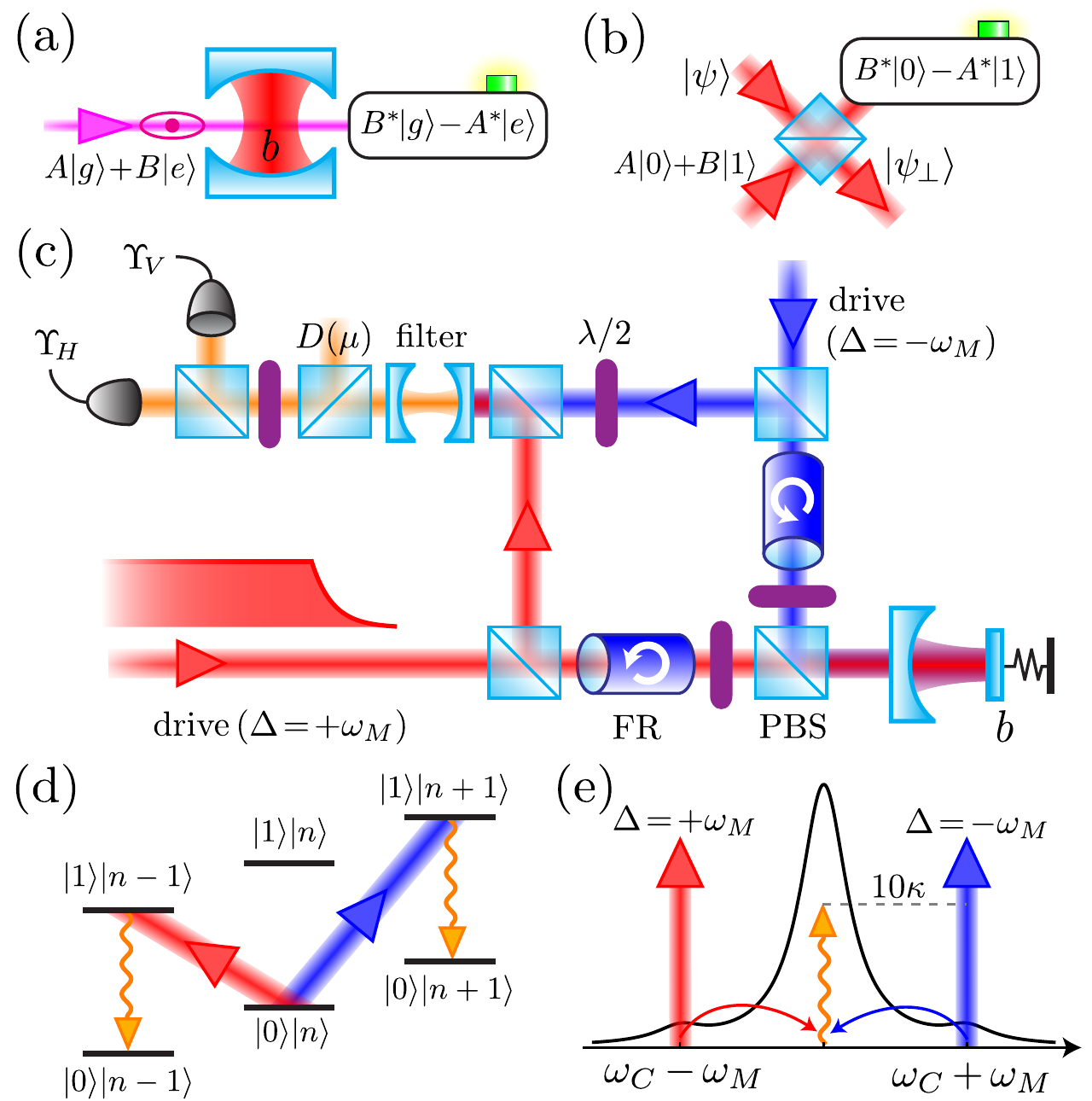}
\caption{
A continuous variable pure state can be orthogonalized by coupling with a qubit via the Jaynes-Cummings (a) or the beam-splitter (b) interaction and then measurement of the qubit. Alternatively, simultaneously using the beam-splitter and two-mode-squeezing interactions can be used for state orthogonalization. This can be realized with cavity optomechanics to coherently manipulate the quantum state of motion of a mechanical oscillator (c). (PBS: polarizing beam splitter, FR: Faraday rotator). One of the drive fields is blue detuned and gives rise to a phonon-number-increasing process whereas the other is red detuned and gives rise to a phonon-number-reducing process. This is shown in (d), a truncated energy level diagram of the optomechanical system where the left kets describe the intracavity photon number and the right kets describe the mechanical phonon number. Each drive generates a sideband at cavity resonance, which is shown in (e), an optomechanical spectrum. Thus, after erasure of the polarization information, photon detection at this frequency causes the mechanical element to undergo a coherent superposition of phonon addition and subtraction.
}
\label{Fig:Scheme}
\end{figure}



Our proposed setup for coherent phonon control uses two orthogonally polarized optical fields to interact with the mechanical resonator, see Fig.~1(c). We consider a pulsed protocol where the conditional mechanical state following the pulsed interaction and measurement is determined. The optomechanical Hamiltonian~\cite{Law1995} for the two independent optical modes in the optical rotating frame at the drive frequencies is
\begin{equation}
\begin{split}
\frac{H}{\hbar} = \omegam b^\dagger b + \!
\sum_{i} \! \left( \detune{i} a^\dagger_i a_i - g_0 a^\dagger_i a_i (b+b^\dagger) \right) + \frac{H_d}{\hbar},
\end{split}
\end{equation}
where $H_d/\hbar = \sum_i \sqrt{2\kappa N_i}(\mathcal{E}_i^* a_i + \mathcal{E}_i a^\dagger_i)$ is the drive term, the subscripts label the two orthogonally polarized modes $i \jj{\in} \{h,v\}$, and $a$ ($b$) is the cavity (mechanical) annihilation operator. ($\omegam$, mechanical angular frequency; $\Delta$, optical detuning; $g_0$, optomechanical coupling rate; $\kappa$, cavity amplitude decay rate; $N$, photon number per pulse; $\mathcal{E}$, drive amplitude, where $\integ{t}\!\abs{\mathcal{E}}^2 \jj{=} 1$.) Neglecting mechanical damping and input noise, as the interaction time can be made shorter than the decoherence time scale, we compute the dynamics in a similar manner to Ref.~\cite{Zhang2003}. The mechanical evolution is computed via the Hamiltonian and the cavity field is computed via the Langevin equation $\dot{a}_i = -ia_i\left[\Delta_i~-~g_0(b+b^\dagger)\right] + \sqrt{2\kappa}(a_{\textrm{in},i}~-~i\sqrt{N_i}\mathcal{E}_i) - \kappa a_i$, where $\ai$ is the optical input noise. We enter a displaced frame to follow the mean of the operators, i.e. $a_i \rightarrow \sqrt{N_i}\alpha_i + a_i$ and $b \rightarrow \beta + b$. Provided that the intracavity intensity varies much slower than the mechanical frequency the mechanical mean amplitude is $\beta \simeq \frac{g_0}{\omegam}\sum_i N_i \! \abs{\alpha_i}^2$. This displacement due to the optical steady state intensity shifts the mean cavity length. Introducing $\Delta'_i = \Delta_i - 2g_0\beta$, the intracavity amplitude is $\alpha_i \jj{\simeq} -i\sqrt{2\kappa}\mathcal{E}_i/(i\Delta'_i + \kappa)$, where it has been assumed that $\mathcal{E}$ varies much slower than $\kappa$. In the proceeding discussion this change to the detuning is neglected as the effect is small and can be readily compensated by frequency stabilization and/or appropriate pre-detuning. We turn now to the noise operators and for brevity solve the dynamics for a single drive frequency. We enter the mechanical and optical rotating frames via $a \rightarrow a e^{-i\Delta t}$ and $b \rightarrow b e^{-i\omegam t}$, respectively. Assuming $\kappa \ll \omegam$, we make the rotating-wave approximation and obtain $\dot{a} = i g_0 \sqrt{N} \alpha \, b^{(,\dagger)} + \sqrt{2\kappa} \ai - \kappa a$ and $\dot{b} = i g_0 \sqrt{N} \alpha^{(*,)} a^{(,\dagger)}$, where the brackets in the superscripts are used to describe the two detunings we consider $(\Delta = +\omegam, \Delta = -\omegam)$ respectively. For $g_0 \sqrt{N} \alpha \ll \kappa$ we use the adiabatic solution $a \simeq i \frac{g_0}{\kappa} \sqrt{N} \alpha b^{(,\dagger)} + \zeta$, where $\zeta(t) = \sqrt{2\kappa}\integlim{-\infty}{t}{t'}e^{-\kappa(t-t')}\ai(t')$. The photon number scattered by the optomechanical interaction is $n = \integlim{0}{\tau}{t}\aod \ao$, which has been approximated to include detection up to the drive duration $\tau \jj{\gg} \kappa^{-1}$ and $\ao \jj{=} \sqrt{2\kappa}a - \ai$ is the cavity output. For the $h$ polarization driving the beam-splitter interaction ($\Delta\jj{=}{+\omegam}$), $\mean{n_h} = (1 - e^{-2G_h\tau})\langle\bzerod \bzero\rangle$, where $G_i = \frac{g_0^2}{\kappa}N_i\abs{\alpha_i}^2$ and $\bzero$ is the mechanical field operator at the beginning of the interaction, time $t\jj{=}0$. For the $v$ polarization ($\Delta\jj{=}{-\omegam}$), which drives the two-mode-squeezing interaction, $\mean{n_v} \jj{=} (e^{2G_v\tau} - 1)\langle\bzero \bzerod\rangle$. We now consider weak drive such that the probability of more than one quanta being scattered is negligible. In this case, from the scattered photon number expectations, we introduce an effective beam-splitter parameter $\bsParam \jj{=} \sqrt{2G_h\tau}$ and an effective squeezing parameter $r \jj{=} \sqrt{2G_v\tau}$~\cite{Scully1997} and we describe the interaction using the effective unitary $\Ueff = 1 + (\bsParam a_h^\dagger b e^{-i\phi} - r a_v^\dagger b^\dagger e^{i\varphi} - \hc)$ \cite{note:Supplementary}. Here, $\phi$ and $\varphi$ are the beam-splitter and two-mode-squeezer phases, respectively, which can be controlled via the phase of the drives. The fields at cavity resonance generated via $\Ueff$ are spatially combined and filtered from the drive fields. Next, to control the weighting of identity in the operation a weak displacement of amplitude $\mu$ is performed \cite{ref:NotePegg1998}. Doing this to the $h$ polarization, $\Ueff \rightarrow 1 + (\bsParam a_h^\dagger b e^{-i\phi} - r a_v^\dagger b^\dagger e^{i\varphi} + \mu a_h^\dagger - \hc)$. At this point the polarization of a scattered photon reveals how the phonon number changed. The field then passes through a wave-plate that performs $a_h \rightarrow \frac{1}{\sqrt{2}}(a_h + a_v)$ and $a_v \rightarrow \frac{1}{\sqrt{2}}(a_v - a_h)$ and is then incident upon a polarizing beam splitter to conceal this information and allow for a quantum superposition. Conditioned on a $h$ photon detection, the resulting mechanical state is $\rhoMout = \UpsH \rhoMin \UpsHd / \prob{h}$, where $\prob{h} = \traceSubInline{M}{\UpsHd\UpsH \rhoM}$ is the probability of photon detection and
\begin{equation}
\label{eq:UpsH}
\UpsH = \frac{1}{\sqrt{2}}\left( \bsParam b e^{-i\phi} + r b^\dagger e^{i\varphi} + \mu \right).
\end{equation}
%
A $v$ photon detection gives a measurement operator of the same form, however, with a $\pi$ phase shift on the identity.



\emph{Applications.---} $\UpsH$ provides a method to prepare and manipulate quantum coherence between the mechanical energy levels. Setting $\mu\jj{=}0$, $\bsParam\jj{=}r$ and $\phi\jj{=}\varphi\jj{=}\vartheta\jj{+}\pi/2$ we obtain the \emph{quantum state orthogonalizer} $\UpsPerp = r(b e^{-i(\vartheta + \pi/2)} + b^\dagger e^{i(\vartheta + \pi/2)})/\sqrt{2} = r \PM^{(\vartheta)}$. This quadrature is depicted in Fig.~2 as is its action on a displaced squeezed state. Such orthogonalization is heralded by the detection of a single photon that occurs with probability $\prob{h} = r^2 \mean{(\PM^{(\vartheta)})^2}$, which is greater than zero for all physical states \cite{note:QuadEigenstate}. We also note here that for states with zero phase-space mean, i.e. $\bra{\psi}b\ket{\psi} = \bra{\psi}b^\dagger\ket{\psi} = 0$, one can also interpret these expressions as quanta subtraction or addition to the state $\ket{\psi}$ yields a state which is orthogonal to $\ket{\psi}$. Addition alone can orthogonalize all such states with a heralding probability of $r^2(\mean{b^\dagger b}+1)/2$ whereas subtraction alone has a heralding probability of $(\bsParam)^2\mean{b^\dagger b}/2$. We thus further note that the operations $b\jj{-}\beta$ and $b^\dagger\jj{-}\beta^*$ can orthogonalize all pure states with $\mean{b}\jj{=}\beta$. These operations may be simpler to experimentally implement, however, are less versatile as complete information of the state's mean is required as opposed to the partial knowledge required by $\UpsPerp$. Returning to (\ref{eq:UpsH}) one can now form a superposition of orthogonalization and identity, $\UpsH\jj{=}\mu/\sqrt{2} + \Upsilon_{\!\perp}$, to prepare a superposition of the initial state and an orthogonal state, i.e. a qubit, see Fig.~2(d).

\begin{figure}[t!]
\includegraphics[width=0.9\hsize]{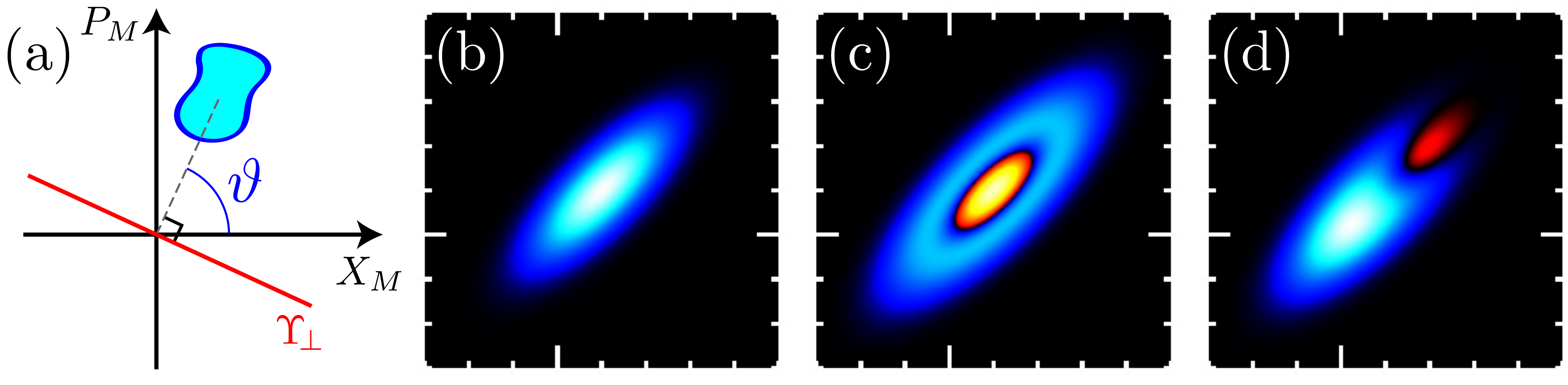}
\caption{
An equally weighted superposition of quanta addition and subtraction can orthogonalize any pure quantum state. (a) The orthogonalizer $\Upsilon_{\!\perp}$ is a quadrature perpendicular to the angle $\vartheta$ made by the input state's mean in phase space. The Wigner function (blue-cyan: positive, red-yellow: negative, larger ticks mark the origin and they increment by unity) of a displaced squeezed state (b), which has been orthogonalized (c). 
A superposition of an initial state with an orthogonal state may be prepared to create a qubit from any initial pure state. In (d) such a superposition is shown by action with $\Upsilon_{\!\perp} + \abs{\mu} e^{-i\pi/2}/\sqrt{2}$, where $\abs{\mu} = r$.
}
\label{Fig:Orthog}
\end{figure}


A mechanical resonator is a stationary system that allows $\UpsH$ to be conveniently performed multiple times. Moreover, as the superposition weightings can be changed between applications this provides considerable versatility for quantum state engineering and quantum control protocols. For instance, one could realize the protocol by Dakna \etal~\cite{Dakna1999} to synthesize an arbitrary mechanical motional state. As another application, here we show that with $N$ applications of $\UpsH$, one can transform the state $\ket{\psi}\jj{=}\sum_n^N \psi_n \ket{n}$ into any target state $\ket{\phi}\jj{=}\sum_n^N \phi_n \ket{n}$, i.e. \emph{arbitrary quantum state transformation}. Our method uses only the subtraction and identity components of $\UpsH$ \cite{note:coherentState} and proceeds in a manner similar to Ref.~\cite{Dakna1999} and generalizes the scheme presented in Ref.~\cite{Fiurasek2005}. Specifically, by applying $\Phi = \prod_{j=1}^N (\mu_j + \nu_j b)/\sqrt{2} = \sum_{i=0}^N C_i b^i$, where $\nu = \bsParam e^{-i\phi}$, to the state $\ket{\psi}$ one can obtain $\ket{\phi}$ provided that the set of coefficients $C_i$ is such that $\sum_{i=0}^{N-n} C_i \, \psi_{i+n} \sqrt{(i+n)!/n!} = \phi_n$. Determining $C_i$ can be readily achieved via matrix inversion and a solution exists provided that $\psi_N\jj{\neq}0$ \cite{note:Supplementary}. For the initial state having $\mean{b} = 0$, the probability of successful quantum state transformation is $\prod_{i=1}^N [(\bsParam)_i^2\mean{b^\dagger b}_i + \abs{\mu_i}^2]/2$, where $\mean{b^\dagger b}_i$ is the phonon number expectation prior to the $i$th pulse. This probability may seem low, however, the experiment can readily be performed with a megahertz repetition rate using a $\sim$~100~MHz mechanical oscillator and thus a practical number of heralding events can be attained in a reasonable time. Also, if the target state has a larger (smaller) dimension than the initial state one can apply creation (annihilation) as many times as necessary in order to make the dimensions the same prior to using $\Phi$.

\emph{An experimental approach.---} There are numerous realizations of optomechanical systems and much progress has been made that can be built upon; the most pertinent being  Refs. \cite{Lee2011, Lee2012} where phonon addition and subtraction were realized as separate operations. Combining these operations into a coherent superposition can be achieved with the setup in Fig.~1(c). Here we present an alternative route to fulfill the requirements of our proposed scheme using a mechanical element with a bulk acoustic wave vibration that forms an end mirror of a Fabry-P\'{e}rot cavity~\cite{Kuhn2011}. This configuration has the advantage that the cavity decay can be controlled independently of the mechanical properties and such vibrational modes offer high mechanical resonance frequencies \cite{Borkje2012}. Moreover, simultaneous high reflectivity and high mechanical quality can be realized with multilayer crystalline reflectors~\cite{Cole2008}. A 40~$\mu$m diameter and several micrometer thick mirror has a mechanical resonance $\omegam/2\pi \jj{=} 200$~MHz with a 20~ng effective mass. With a finesse of $5\jj{\times}10^4$, to achieve resolved sideband operation, i.e. $\omegam/\kappa \jj{=} 10$, a 75~$\mu$m cavity length can be used. For a drive laser with wavelength 1064~nm and a pulse duration of one hundred mechanical periods an optical power during the pulse of 1.3~mW is needed to achieve $r^2 \jj{=} 0.01$. During the interaction the mechanical resonator also interacts with its thermal environment. To neglect the effects of environmental coupling we require that $\xi = (\nbar / Q) (\tau \omegam / 2\pi) \ll 1$, where $\nbar$ is the mechanical phonon occupation in thermal equilibrium and $Q$ is the mechanical quality factor. For $Q \jj{=} 10^5$ and a $100$~mK bath, which can be readily achieved using dilution refrigeration, $\xi \jj{\simeq} 10^{-2}$. Following the interaction the sideband needs to be separated from the drive field(s) prior to photon detection. For higher mechanical frequencies the filtering requirements simplify. However, it is possible to achieve sufficient filtering even for a 200~MHz mechanical frequency using an optical displacement and spectral filtering \cite{note:filtering}. Realizing the displacement with optical-fiber-based components, which provide excellent spatial mode matching, one can achieve an interferometric visibility of $99.99\%$ that suppresses the drive by $10^{4}$. The remaining drive can be further reduced by filtering with a cavity that has the same resonance frequency as the optomechanical cavity. To achieve a drive transmission $10^2$ times smaller than sideband transmission, a filter cavity amplitude decay rate of 2~kHz is required~\cite{note:cavity}. We would also like to emphasize that our scheme is robust against optical loss and inefficient detection as an optomechanically scattered photon that goes undetected  does not trigger $\UpsH$, hence, the primary effect of loss is to merely reduce the heralding probability~\cite{note:twoPhoton}. To characterize the mechanical motional state, as the parameter regime considered here is suited for the beam-splitter interaction, quantum state transfer of the mechanical motional state to the light~\cite{Parkins1999,Zhang2003} can be performed followed by optical homodyne tomography. This interaction, following action(s) with $\UpsH$ to the stationary mechanical element, also provides a route to prepare optical continuous variable qubits or to synthesize arbitrary quantum states of a travelling optical field.



\emph{Conclusions.---} A superposition of quanta addition and subtraction can orthogonalize any pure continuous-variable quantum state with known angle made by the mean of the state's amplitude in phase-space. Such a superposition in combination with a controllable amount of the identity operation provides extensive control for quantum state engineering and quantum information applications. For stationary systems it is convenient to apply this tool multiple times, which we have utilized to illustrate how to perform arbitrary quantum state transformation. As the interactions we have used are available in many of the facets of quantum optics~\cite{Hammerer2010}, the tools we introduce can be realized in numerous physical systems.


\emph{Acknowledgments.---}
We thank G.~D.~Cole, S.~G.~Hofer and K.~Hammerer for useful discussion. M.R.V. is a member of the FWF Doctoral Programme CoQuS (W 1210), is a recipient of a DOC fellowship of the Austrian Academy of Sciences and gratefully acknowledges the Royal Society. We thank support provided by the European Comission (Q-ESSENCE), the European Research Council (ERC QOM), the Austrian Science Fund (FWF) (START, SFB FOQUS), the Foundational Questions Institute, the John Templeton Foundation and the Qatar National Research Fund (NPRP 4-554-1-084).


\newpage

\appendix

\section*{Supplementary Material}

\subsection{Determining the Effective Unitary Interaction}

Central to our discussion in the main text is the measurement operator $\UpsH$, which is used to describe the operation to the mechanical resonator via the optomechanical interaction and then single photon detection. $\UpsH = \bra{1,0}\Ueff\ket{0,0}$, where the ket is the initial state of light at the cavity resonance for the two orthogonal polarizations used, the bra describes a $h$-polarization photon detection with no $v$ photon detection, and $\Ueff$ is the effective optomechanical interaction including the manipulations to the optical field made after interaction with the mechanical resonator. In this supplementary we provide a discussion how $\Ueff$ is obtained.

The time evolutions described in Eq.~(2) of the main text are generated by the beam-splitter and two-mode-squeezing effective interaction Hamiltonians. In the former case $a$ accumulates correlation with $b$ and in the latter case $a$ accumulates correlation with $b^\dagger$. For vacuum on the input of mode $a$, the expectation of the number operator in the output of mode $a$ for the beam-splitter and two-mode-squeezing interactions are
\begin{displaymath}
\sin^2\! \bsParam \, \mean{b^\dagger b}, \quad \textrm{and} \quad
\sinh^2\! r\, \mean{b b^\dagger},
\end{displaymath}
respectively, where $\sin^2(\bsParam)$ is the (intensity) reflectivity of the beam-splitter and $r$ is the squeezing parameter. In the optomechanical scheme we have considered, the mean photon number scattered by the optomechanical interaction for the beam-splitter and two-mode-squeezing interactions are
\begin{displaymath}
\mean{n_h} = (1 - e^{-2G_h\tau})\langle\bzerod \bzero\rangle, \,\, \textrm{and} \,\,
\mean{n_v} = (e^{2G_v\tau} - 1)\langle\bzero \bzerod\rangle,
\end{displaymath}
respectively. For small $\bsParam$, $r$, and $G\tau$ we then have
\begin{displaymath}
\bsParam = \sqrt{2G_h\tau}, \quad \textrm{and} \quad
r = \sqrt{2G_v\tau},
\end{displaymath}
for the effective optomechanical beam-splitter and two-mode-squeezing parameters, respectively. It is noted here that computing the mean number output in mode $b$ can also be performed to yield these parameters. As both the beam-splitter and two-mode-squeezing processes are driven simultaneously, we expect that the effective optomechanical unitary take the form $\Ueff = \exp \left[-\frac{i}{\hbar}(H_{BS} + H_{SQ})\tau\right]$, where $H_{BS} \propto a^\dagger b + a b^\dagger$ and $H_{SQ} \propto ab + a^\dagger b^\dagger$ are the beam-splitter and two-mode-squeezing Hamiltonians respectively. To first order in the beam-splitter and squeezing parameters the effective unitary describing the cavity optomechanical interaction is then
\begin{displaymath}
\Ueff = 1 + (\bsParam a_h^\dagger b e^{-i\phi} - r a_v^\dagger b^\dagger e^{i\varphi} - \hc).
\end{displaymath}
Finally, to obtain the effective unitary used for the measurement operator, the polarization manipulations to the optical fields, as discussed in the main text,  must be performed.


\subsection{Arbitrary Quantum State Transformation}
In the main text we introduced a scheme for \emph{arbitrary quantum state transformation} that generates a target state from a known input state. Here we further discuss our protocol and provide a specific quantum state transformation example.

The protocol works as follows. For a known initial state
\begin{displaymath}
\ket{\psi} = \sum_{n=0}^N \psi_n \ket{n},
\end{displaymath}
which has no excitation beyond $N$ quanta (or has been approximated by truncation at this level), any target state of the form
\begin{displaymath}
\ket{\phi} = \sum_{n=0}^N \phi_n \ket{n},
\end{displaymath}
can be generated by applying a controllably weighted superposition of identity and subtraction $N$ times, i.e.
\begin{equation}
\Phi = \prod_{j=1}^N (\mu_j + \nu_j b)/\sqrt{2} = \sum_{i=0}^N C_i b^i.
\label{eq:munuC}
\end{equation}
Applying this operation to the initial state we have
\begin{displaymath}
\Phi \ket{\psi} = \sum_{i=0}^N \sum_{k=0}^N C_i \psi_k \sqrt{\frac{k!}{(k-i)!}}\ket{k-i},
\end{displaymath}
where we have used $b\ket{n} = \sqrt{n}\ket{n-1}$. 

The operation $\Phi$ is a non-unitary process and the un-normalized matrix elements of the state after application of $\Phi$ are
\begin{equation}
\label{eq:Suppphin}
\bra{n}\Phi\ket{\psi} = \sum_{i=0}^{N-n} C_i \, \psi_{i+n} \sqrt{\frac{(i+n)!}{n!}}.
\end{equation}
The target state $\ket{\phi}$ is reached when $\bra{n}\Phi\ket{\psi} = \phi_n$. Provided that $\psi_N \neq 0$ a set of coefficients $C_i$ fulfilling $\bra{n}\Phi\ket{\psi} = \phi_n$ can be determined via matrix inversion. Once a set of coefficients $C_i$ is determined, a set of complex coefficients $\mu_j$ and $\nu_j$ that satisfy (\ref{eq:munuC}) can also readily be determined via matrix inversion.

\begin{widetext}

\begin{figure*}[h!]
\includegraphics[width=1.0\hsize]{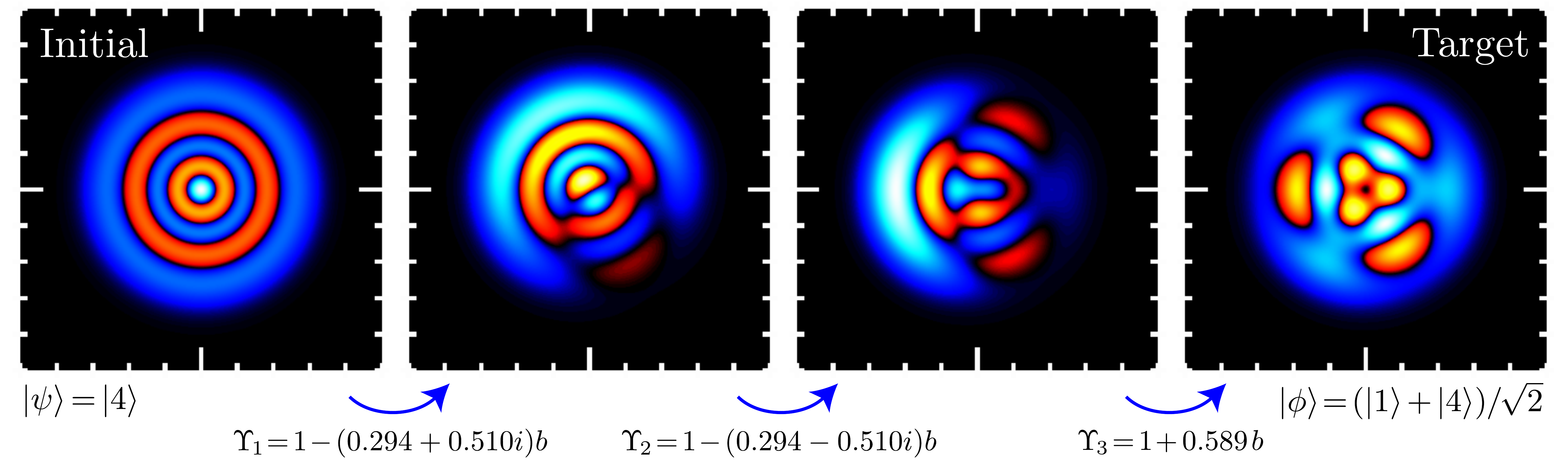}
\caption{
An example quantum state transformation. Shown are Wigner functions (blue-cyan: positive, red-yellow: negative, larger ticks mark the origin and they increment by unity) of an initial Fock state (left) to a target state (right). The target state is reached by a sequence of three operations of a controllably weighted superposition of identity and subtraction. The relative amplitude between identity and subtraction for each step is shown.
}
\label{Fig:TransformExample}
\end{figure*}

\end{widetext}

We now provide a specific example of a quantum state transformation. Starting with an initial state $\ket{\psi} = \ket{4}$ we wish to reach the target state $\ket{\phi} = (\ket{1} + \ket{4})/\sqrt{2}$. This target state can be reached with three applications of identity and subtraction. Solving (\ref{eq:Suppphin}) we find that $C_0~=~\sqrt{24} C_3$ and $C_1 = C_2 = 0$. As identity has been used with each application we set $\mu = 1$ and obtain
\begin{align}
\nu_1 + \nu_2 + \nu_3 &= 0, \nonumber\\
\nu_1 \nu_2 + \nu_1 \nu_3 + \nu_2 \nu_3  &= 0, \nonumber\\
\nu_1 \nu_2 \nu_3 \sqrt{24} &= 1. \nonumber\\
 \nonumber
\end{align}
These equations can be readily solved exactly to provide the relative amplitudes between identity and subtraction to produce the target state. Numerical approximations to the solutions and the intermediate states during the quantum state transformation process are shown in Fig.~\ref{Fig:TransformExample}.

\begin{thebibliography}{99}
\bibitem{note:SpinHalfAnalog} It is interesting to note that this method may also be adapted to spin-1/2 systems using a superposition of the raising and lowering operators, however, in this case one requires knowledge of the phase between the two spin states thus rendering orthogonalization trivial.
\bibitem{note:QuadEigenstate} A quadrature eigenstate can have $\mean{(\PM^{(\vartheta)})^2} = 0$, however, this is not a physical state.
\bibitem{note:cavity} A ring cavity with a 4 m cavity perimeter can provide this required linewidth with the modest finesse of $2\times 10^4$. Alternatively, two (or more) larger linewidth cavities can be used in series to provide the required filtering.
\bibitem{note:twoPhoton} For a given detection event, however, there is the possibility that a second photon was missed due to optical loss, which was due to a two-phonon operation. This poses no problem to our scheme as the drive fields are chosen here such that the probability of two-photon scattering is negligible.
\end{thebibliography}
\end{document}